\documentclass{article}
\usepackage{nopageno}
\usepackage[utf8]{inputenc}
\usepackage{graphicx}
\usepackage{amsmath,amssymb}
\usepackage{afterpage}
\usepackage{enumitem}
\usepackage{esvect}
\usepackage{algorithm}
\usepackage[noend]{algpseudocode}
\usepackage{hyperref}
\usepackage{amsfonts}
\usepackage{amssymb,amsmath,amsthm,bm,bbm}
\usepackage{multirow}
\usepackage[font=small,labelfont=bf,tableposition=top]{caption}
\usepackage[table]{xcolor}
\DeclareCaptionLabelFormat{andtable}{#1~#2  \&  \tablename~\thetable}
\usepackage{graphicx}
\usepackage{subfig}
\usepackage{tikz}
\usepackage{authblk}
\usetikzlibrary{positioning}
\title{A Recurrent Neural Network and Differential Equation Based Spatiotemporal Infectious Disease Model with Application to COVID-19} 
\author[1]{Zhijian Li}
\author[1]{Yunling Zheng}
\author[1]{Jack Xin}
\author[2]{Guofa Zhou}
\affil[1]{Department of Mathematics, UC Irvine}
\affil[2]{College of Health Sciences, UC Irvine,
CA 92697, USA.}
\date{July 21, 2020.}
\begin{document}

\maketitle


\abstract{The outbreaks of Coronavirus Disease 2019 (COVID-19) have impacted the world significantly. Modeling the trend of infection and real-time forecasting of cases can help decision making and control of the disease spread. 
However, data-driven methods such as recurrent neural networks (RNN) can perform poorly due to limited daily samples in time. In this work, we develop an integrated spatiotemporal model based on the epidemic differential equations (SIR) and RNN. The former after simplification and discretization is a compact model of temporal infection trend of a region while the latter models the effect of nearest neighboring regions. The latter captures latent spatial  information.
We trained and tested our model on COVID-19 data in Italy, and show that it   out-performs existing temporal models (fully connected NN, SIR, ARIMA) in 1-day, 3-day, and  1-week ahead  forecasting especially in the  regime of limited training data.}
\medskip


\hspace{.5 in} 
Discrete Epidemic Model,
Spatiotemporal Machine Learning.

\onecolumn \maketitle \normalsize \setcounter{footnote}{0} \vfill

\section{Introduction}
\label{sec:introduction}

\noindent Susceptible-Infected-Removed (SIR)
is a classical  
differential equation 
model of infectious diseases \cite{anderson_may_1992}. It divides the total population into three compartments and models their evolution by the system of equations
$$\frac{dS}{dt}=\, -\beta \, I\, S$$
$$\frac{dI}{dt}=\, \beta \, I\, S-\gamma\, I$$
$$\frac{dR}{dt}=\, \gamma \, I$$
where $\beta$ and $\gamma$ are two  positive parameters. SIR is a simple and efficient model of temporal data for a given region, see also  \cite{hethcote_2000} for related compartment models with social structures.
 \medskip
 
Yet the infectious disease data are often spatio-temporal as in 
the case of COVID-19, see \cite{Italy_data}. 
A natural question is how to extend SIR to a space time model of suitable complexity so that it can be quickly trained from the available public data sets and applied in real-time forecasts. See
\cite{roosa_2020} for temporal model real-time  forecasts on cumulative cases of China in Feb 2020. 
\medskip

In this paper, we explore spatial infectious disease information to model the latent effect due to the in-flow of the infected people from the  geographical neighbors. The in-flow data is not observed. To this end, machine learning tools such as regression and neural network models are more convenient.
Auto-regressive model (AR) and its variants are linear statistical models to forecast time-series data. 
The Long Short Term Memory (LSTM) neural networks, originally designed for natural language processing \cite{hochreiter1997long}, have more representation power and can be applied to disease time-series data as well. With spatial structures added, 
the graph-structured LSTM models can achieve state-of-the-art performance on spatiotemporal  influenza data \cite{li_2019}, crime and traffic data \cite{wang,wang2018graph}. 
However, they require a large enough supply of training data. 
For COVID-19, we only have limited daily data since the outbreaks began in early 2020. Applying space-time LSTM models \cite{li_2019,wang2018graph} directly to COVID-19 turns out to produce poor results.   
In view of the limited COVID-19 data, we shall  propose a hybrid SIR-LSTM model.

\section{Related Work}
\noindent In \cite{yang2015accurate}, the authors  designed a variant of AR, the AutoRegression with Google search data (ARGO), that utilizes external feature of google search data to forecast influenza data from Centers for Disease Control and Prevention (CDC). Based on google search trend data that correlated to influenza as external feature, ARGO is a linear model that processes  historical observations and external features. The prediction of influenza activity level at time $t$, defined as $\hat{y}_t$, is given by: 
$$\hat{y}_t=u_t+\sum_{j=1}^{52}\alpha_jy_{t-j}+\sum_{i=1}^{100}\beta_iX_{i,t}.$$
ARGO is optimized as:
$$\min_{\mu_y,\vec{\alpha},\vec{\beta}}\Big(y_t-u_t-\sum_{j=1}^{52}\alpha_jy_{t-j}-\sum_{i=1}^{100}\beta_iX_{i,t}\Big)^2$$
$$+\lambda_a||\vec{\alpha}||_1+\eta_a||\vec{\beta}||_1+\lambda_b||\vec{\alpha}||_2^2+\eta_b||\vec{\beta}||_2^2$$
where $\vec{\alpha}=(\alpha_1,\cdots,\alpha_{52})$ and $\vec{\beta}=(\beta_1,\cdots,\beta_{100})$. The $y_{t-j}$'s, $ 1\leq j\leq 52$, are historical observations of previous 52 weeks and $X_{i,t}$ are the google search trend measures of top 100 terms that are most correlated to influenza at time $t$. 
Essentially, ARGO is a linear regression with regularization terms. In \cite{yang2015accurate}, ARGO is shown to outperform standard machine learning models such as LSTM, AR, and ARIMA.
\medskip

In \cite{li_2019},  graph  structured recurrent neural network (GSRNN) further improved ARGO in the forecasting accuracy of CDC influenza  activity level. The CDC partitions the US into 10 Health and Human
Services (HHS) regions for reporting. GSRNN treats the 10 regions as a graph with nodes $v_1,\cdots,v_{10}$, and $E$ be the collection of edges (i.e $E=\{(v_i,v_j)|v_i, v_j \,\,   \text{are adjacent}\}$). Based on the average history of activity levels, the 10 HHS regions are divided into two groups, relatively active group, $\mathcal{H}$, and relatively inactive group, $\mathcal{L}$. There are 3 types of edges, $\mathcal{L}-\mathcal{L}$, $\mathcal{H}-\mathcal{L}$, and $\mathcal{H}-\mathcal{H}$, and each edge type has a corresponding RNN to train the edge features. There are also two node RNNs for each group to output the final prediction. Given a node (region) $v$, suppose $v\in \mathcal{H}$. GSRNN generates the edge features of $v$ at time $t$, $e_{v,\mathcal{H}}^t$ and $e_{v,\mathcal{L}}^t$, by averaging the history of neighbors of $v$ in the corresponding groups. Next, the edge features are fed into the corresponding edge RNNs:
$$f_v^t=\text{edgeRNN}_{\mathcal{H}-\mathcal{L}}(e_{v,\mathcal{L}}^t), \ \  h_v^t=\text{edgeRNN}_{\mathcal{H}-\mathcal{H}}(e_{v,\mathcal{L}}^t)$$
Then, the outputs of edgeRNNs are fed into the nodeRNN of group $\mathcal{H}$ together with the node feature of $v$ at time $t$, denoted as $v^t$, to output the prediction of the activity level of node $v$ at time $t+1$, or $y_{v}^{t+1}$:
$$y_{v}^{t+1}=\text{nodeRNN}_{\mathcal{H}}(v^t,f_v^t,h_v^t).$$
\section{Our Contribution: IeRNN model}
 We propose a novel spatiotemporal model integrating LSTM \cite{hochreiter1997long} with a discrete time I-equation derived from SIR differential equations. The LSTM is utilized to model latent spatial information. The I-equation models the observed temporal information.  Our model, named IeRNN, differs from \cite{li_2019,wang,wang2018graph} in that a difference equation  
 with 3 parameters  (the I-equation) 
 fits the limited temporal data, which is far more compact than LSTM. 

\subsection{Derivation of Discrete-Time  I-Equation}
Based on SIR model, we add an additional feature $I_e$ that represents the external infection influence from the neighbors of a region. Then the SIR nonlinear system with $I_e$ as external forcing becomes
\begin{equation}
\frac{dS}{dt}=\, -\beta_1\, S\, I-\, \beta_2\, S\, I_e
\label{eq:I1}
\end{equation}
\begin{equation}
    \frac{dI}{dt}=\, \beta_1\, S\, I+\, \beta_2\, S\, I_e-\, \gamma \, I
    \label{eq:I2}
\end{equation}
\begin{equation}
    \frac{dR}{dt}=\gamma \, I
    \label{eq:I3}
\end{equation}
which conserves the total mass (normalized to 1): $ S+I + R=1$. 
It follows from (\ref{eq:I3}) that
$$R(t)=R(t_0)+\int_{t_0}^{t}\gamma \, I \, d\tau$$
Hence, 
$$S(t)=1-I(t)-R(t_0)-\, \gamma\, \int_{t_0}^t \, I \,  d\tau$$
Substituting $S(t)$ into (\ref{eq:I2}) we have:
$$\frac{dI}{dt}=(\beta_1\, I+\beta_2\, I_e)\left (1-I(t)-R(t_0)-\gamma \int_{t_0}^tI(\tau) d\tau\right ) - \gamma  I$$
Combining forward Euler method and Riemann sum approximation of the  integral, we have a discrete approximation:
$$I(t+1)=(1-\gamma)I(t)+
\big(\beta_1I(t)  +\beta_2I_e(t)\big)$$ $$\cdot \Big(1-I(t)-R(t_0)-\gamma\frac{t-t_0}{p+1}\sum_{j=0}^{p}I(t-j)\Big)$$
As we model $I(t)$ from the beginning of the infection, we have $t_0=0$ and $R(t_0)=0$. We arrive at the following discrete time I-equation:
\begin{equation}
 I(t+1)=(1-\gamma)I(t)+\big(\beta_1I(t)+\beta_2I_e(t)\big)
    \cdot \Big(1-I(t)-\gamma\frac{t}{p+1}\sum_{j=0}^{p}I(t-j)\Big)
\label{eq:sire}
\end{equation}
\medskip

Note that if we let $I_e(t)=0$, then we have an approximation of $I(t)$ for the original SIR model, which is a solely temporal model (named I model):
\begin{equation}
    I(t+1)= (1-\gamma-\beta)I(t)-\beta\,  I^2(t)
    -\beta \, \gamma\, \frac{t}{p+1}\, I(t)\, 
    \sum_{j=0}^{p}I(t-j) \label{Ieq}
\end{equation}
In reality, it is hard to know how a population of a region interacts with  populations of neighboring regions. As a result, $I_e(t)$ is a latent information that is difficult to model by a mathematical formula or equation. In order to retrieve  latent spatial information, we 
employ recurrent neural networks made of LSTM cells \cite{hochreiter1997long}, see Fig. \ref{fig:lstm}.
\subsection{Generating Edge Feature and Computing Latent $I_e$}
\begin{figure}[ht!]
    \hspace*{-2.5cm}
    \begin{center}
    \includegraphics[scale=0.35]{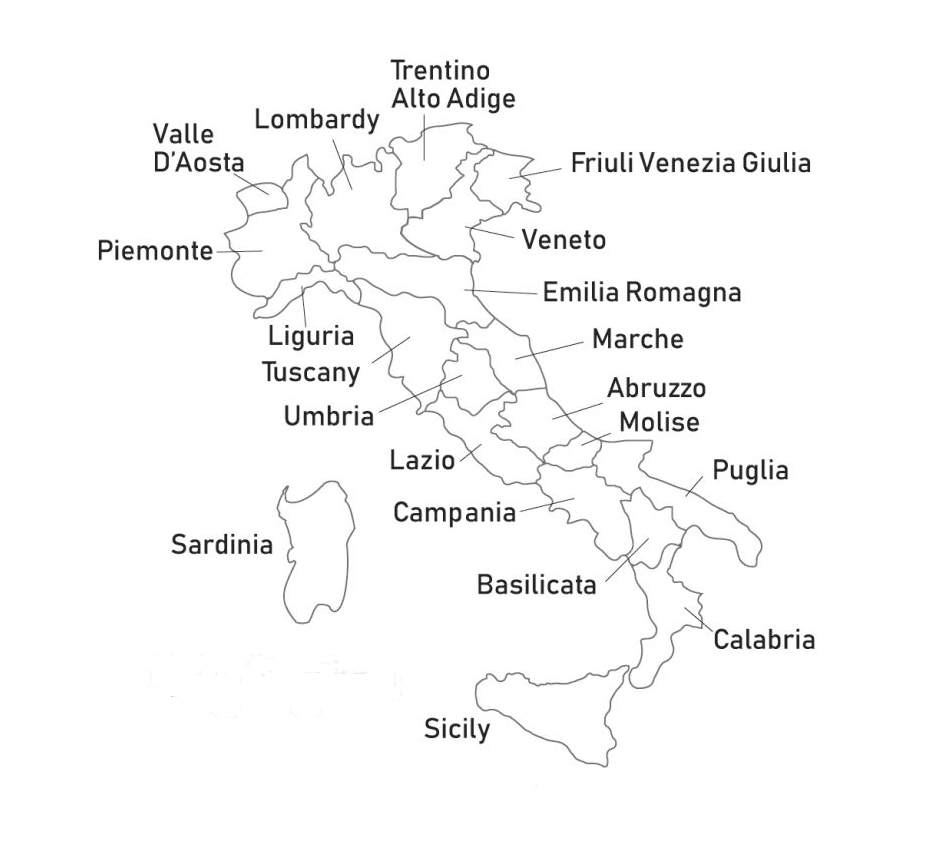}
    \caption{Italian Region Map}
    \label{fig:map}
    \end{center}
\end{figure}

\begin{figure}[ht!]
    \hspace*{-2.5cm}
    \begin{center}
    \includegraphics[scale=0.8]{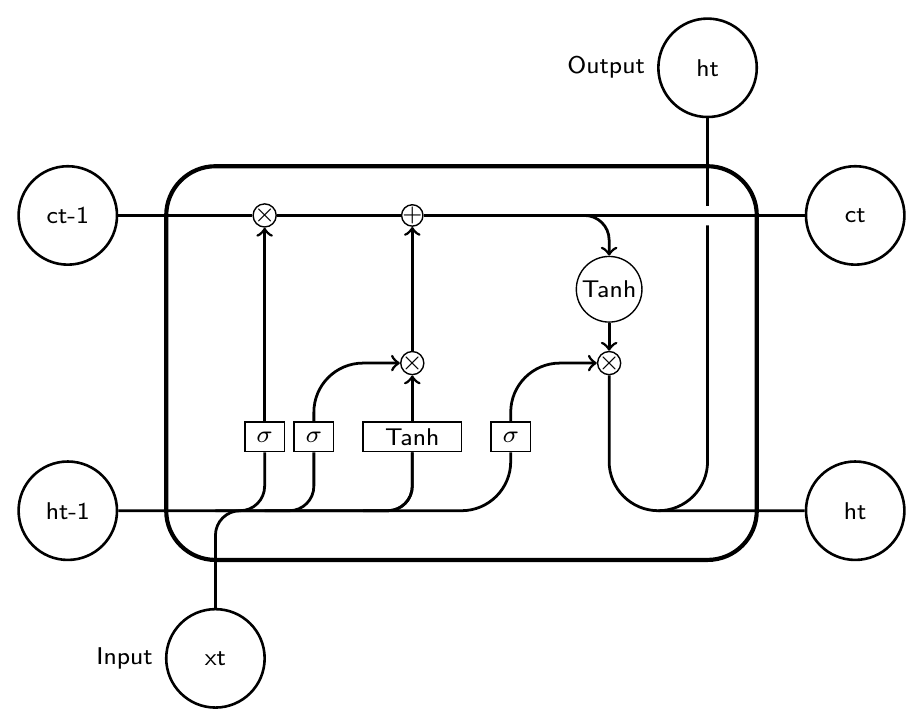}
    \caption{LSTM cell: input $x_t$, output: $h_t$; $\sigma$ is a sigmoid function.}
    \label{fig:lstm}
    \end{center}
\end{figure}

\begin{figure}[ht!]
    \hspace*{-2.5cm}
    \begin{center}
    \includegraphics[scale=0.6]{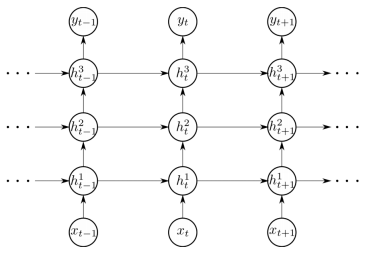}
    \caption{Edge RNN consisting of 3 stacked LSTM cells.}
    \label{fig:ernn}
    \end{center}
\end{figure}

\begin{figure*}[ht!]
    \centering
    \includegraphics[scale=1.7]{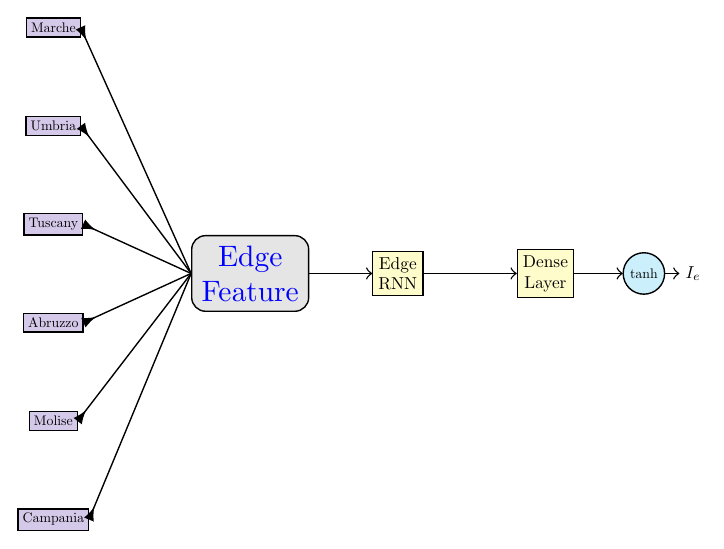}
    \caption{Computing $I_e$ of Lazio region. Edge RNN is as  shown in Fig. \ref{fig:ernn}. Dense layer is fully connected (see Fig. \ref{fig:fcNN}).}
    \label{fig:edge}
\end{figure*}

We utilize the spatial information based on the Italy region map, Fig. \ref{fig:map}. In order to learn the latent information $I_e$ of a region $v$, we first generate the edge feature of $v$. Let $C$ be the collection of neighbors of $v$. Then, the edge feature of $v$ at time $t$ is formulated as:
$$f_e^t=\frac{1}{|C|}\; \sum_{i: \, v_i\in C}\; \big[I_i(t-1),\cdots,I_i(t-p)\big]$$
where $I_i(t)$ is the infection population percentage of region $v_i$ at time $t$.
Then, we feed $f_e^t$ into an Edge RNN, an RNN with 3 stacked LSTM cells (see Fig. \ref{fig:ernn}),  followed by a dense layer for computing $I_e$.  The activation function of the dense layer is 
hyperbolic tangent function. Figure \ref{fig:edge} illustrates the procedure of computing $I_e$ for Lazio as an illustration. We hence {\it call our model IeRNN due to its integrated design of I-equation and edge RNN.} 
\section{Experiment}
To calibrate our model IeRNN, we use the Italy COVID-19 data \cite{Italy_data} for training and testing. Although the US has the most infected cases, the recovered cases are largely missing. On the other hand, the Italian COVID-19 data is more accurately reported and better maintained, reflecting a nearly complete duration of the rise and fall of infection. We collect the data of daily new (current) cases from 2020-02-24 to 2020-06-18 of 20 Italian regions. We set $p=3$ in \eqref{eq:sire} based on experimental performance. As a result, we have the current data for 113 days, with 81 days to train our model and 32 days to test our model (or 70\%/30\% training/testing data split). Our training loss function is the mean squared error of the model output and training data:
$$\hat{y}(t)=(1-\gamma)y(t-1)
+\big(\beta_1y(t-1)+\beta_2I_e(t-1)\big)$$
$$\cdot \Big(1-y(t-1)-\gamma\frac{t}{p+1}\sum_{j=1}^{p+1}y(t-j)\Big)$$
$$Loss=\frac{1}{T-p-1}\sum_{t=p+1}^T\big(y(t)-\hat{y}(t)\big)^2$$
\medskip

{\it Since the training is minimization of the above loss fucntion over parameters in both I-equation and RNN, the two components of IeRNN are coupled while learning from data}. 
We use Adam gradient descent to learn the weights of LSTM and the dense layer, as well as I-equation parameters 
$\beta_1$, $\beta_2$, and $\gamma$.
\medskip

To evaluate the performance of our model, we compare IeRNN, I-model  (\ref{Ieq}), a fully-connected neural network (fcNN, Fig.  \ref{fig:fcNN})  with hyperbolic tangent activation function, and auto-regression model (ARIMA). 
As the standard setting of ARIMA is 1-day ahead prediction, we shall only compare with it in such a very short-term case. Since infectious disease evolution is intrinsically nonlinear, 
we shall compare nonlinear models for 3-day and 1-week 
ahead forecasting. 
 Based on experimental performance, we set the number of hidden units to be 100, 150, and 100 for the three layers of fcNN respectively.

\tikzset{%
  every neuron/.style={
    circle,
    draw,
    minimum size=1cm
  },
  neuron missing/.style={
    draw=none, 
    scale=4,
    text height=0.333cm,
    execute at begin node=\color{black}$\vdots$
  },
}
\begin{figure}[ht!]
    \caption{Schematic of fcNN for modeling time series.}
    \begin{center}
\scalebox{0.6}{%
\begin{tikzpicture}[x=1.5cm, y=1.5cm, >=stealth]
\foreach \m/\l [count=\y] in {1,missing,2}
  \node [every neuron/.try, neuron \m/.try] (input-\m) at (0,2-\y) {};
\foreach \m [count=\y] in {1,missing,2}
  \node [every neuron/.try, neuron \m/.try ] (hidden1-\m) at (1.5,2.5-\y*1.25) {};
\foreach \m [count=\y] in {1,missing,2}
  \node [every neuron/.try, neuron \m/.try ] (hidden2-\m) at (3,2.5-\y*1.25) {};
\foreach \m [count=\y] in {1,missing,2}
  \node [every neuron/.try, neuron \m/.try ] (hidden3-\m) at (4.5,2.5-\y*1.25) {};
\foreach \m [count=\y] in {1}
  \node [every neuron/.try, neuron \m/.try ] (output-\m) at (6,0) {};
\foreach \l [count=\i] in {1,p}
  \draw [<-] (input-\i) -- ++(-1,0)
    node [above, midway] {$\pmb{y_{t-\l}}$};
\foreach \l [count=\i] in {1,n}
  \node [above] at (hidden1-\i.north) {$\pmb{w^{(1)}_\l}$};
\foreach \l [count=\i] in {1,n}
  \node [above] at (hidden2-\i.north) {$\pmb{w^{(2)}_\l}$};
\foreach \l [count=\i] in {1,n}
  \node [above] at (hidden3-\i.north) {$\pmb{w^{(3)}_\l}$};
\foreach \l [count=\i] in {1}
  \node [above] at (output-\i.north) {$\pmb{tanh}$};
\foreach \l [count=\i] in {1}
  \draw [->] (output-\i) -- ++(1,0)
    node [above, midway] {$\pmb{y_t}$};
\foreach \i in {1,2}
  \foreach \j in {1,2}
    \draw [->] (input-\i) -- (hidden1-\j);
\foreach \i in {1,2}
  \foreach \j in {1,2}
    \draw [->] (hidden1-\i) -- (hidden2-\j);
\foreach \i in {1,2}
  \foreach \j in {1,2}
    \draw [->] (hidden2-\i) -- (hidden3-\j);
\foreach \i in {1,2}
  \foreach \j in {1}
    \draw [->] (hidden3-\i) -- (output-\j);
\foreach \l [count=\x from 0] in {\textbf{Input}, \textbf{Hidden}, \textbf{Output}}
  \node [align=center, above] at (\x*3,2.2) {\l \\ \textbf{layer}};
\end{tikzpicture}
}
    \end{center}
    \label{fig:fcNN}
\end{figure}
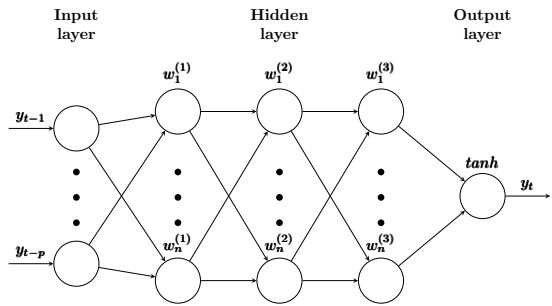

\newpage
\begin{figure*}[ht!]
\caption{Training and 1-day ahead forecast of 4 models (IeRNN, fcNN, I-model, and ARIMA) in 4 rows respectively.}
    \centering
    \subfloat[Lombardy]{\includegraphics[scale=0.45]{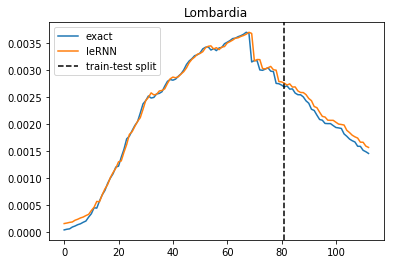}}
    \subfloat[Lazio]{\includegraphics[scale=0.45]{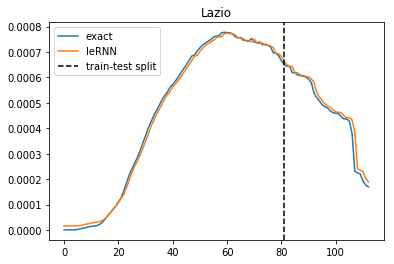}}
    \quad
    \subfloat[Lombardy]{\includegraphics[scale=0.45]{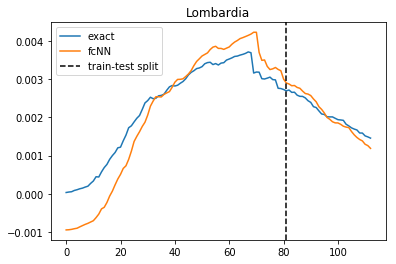}}
    \subfloat[Lazio]{\includegraphics[scale=0.45]{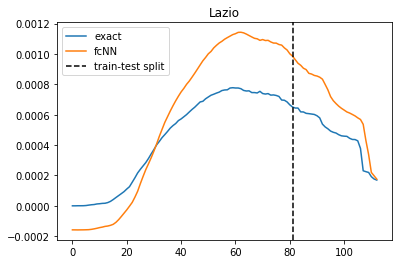}}
    \quad
    \subfloat[Lombardy]{\includegraphics[scale=0.45]{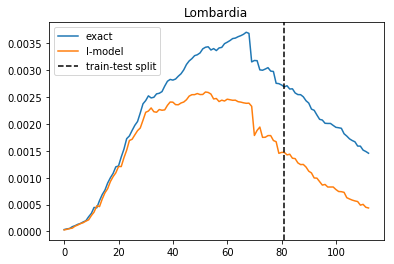}}
    \subfloat[Lazio]{\includegraphics[scale=0.45]{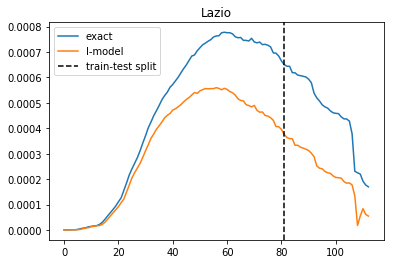}}
    \quad
    \subfloat[Lombardy]{\includegraphics[scale=0.45]{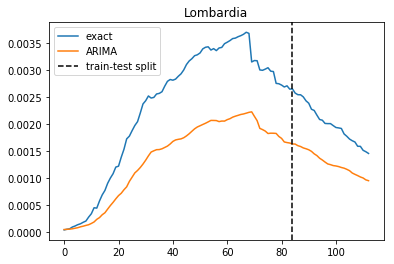}}
    \subfloat[Lazio]{\includegraphics[scale=0.45]{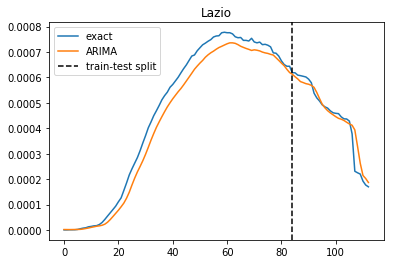}}
    \label{fig:results}
\end{figure*}

\newpage
\begin{figure*}[ht!]
\caption{Training and 
1-day ahead forecast 
of 4 models with reduced (40\%) training data. The 4 rows are IeRNN, fcNN, I-model, and ARIMA respectively.}
    
    \centering
    \subfloat[Lombardy]{\includegraphics[scale=0.45]{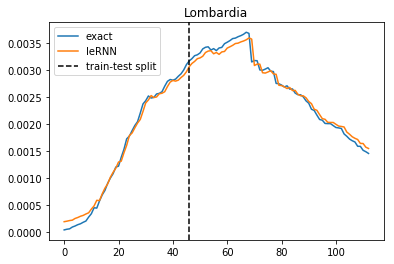}}
    \subfloat[Lazio]{\includegraphics[scale=0.45]{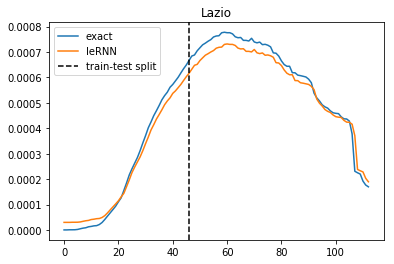}}
    \quad
    \subfloat[Lombardy]{\includegraphics[scale=0.45]{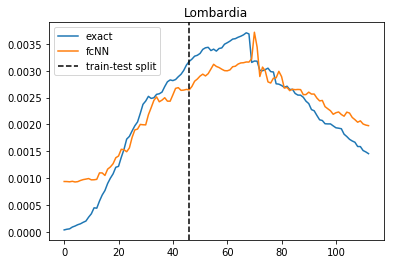}}
    \subfloat[Lazio]{\includegraphics[scale=0.45]{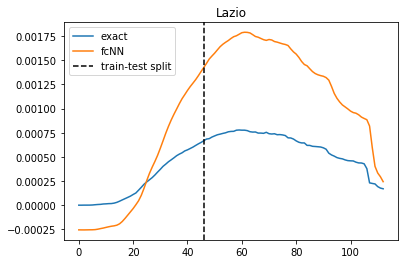}}
    \quad
    \subfloat[Lombardy]{\includegraphics[scale=0.45]{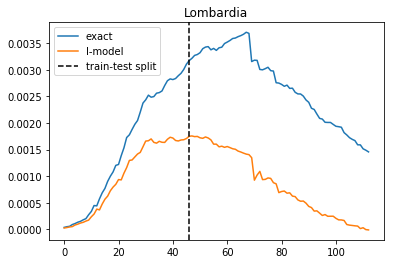}}
   \subfloat[Lazio]{\includegraphics[scale=0.45]{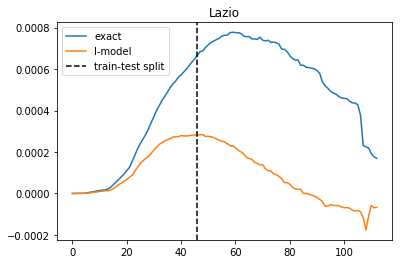}}
    \quad
    \subfloat[Lombardy]{\includegraphics[scale=0.45]{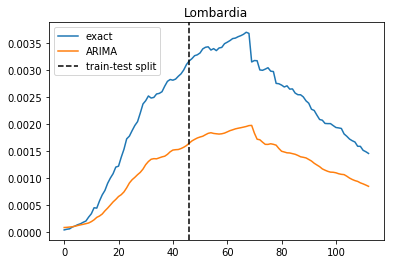}}
    \subfloat[Lazio]{\includegraphics[scale=0.45]{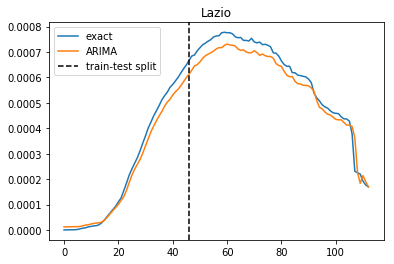}}
    \label{fig:results2}
\end{figure*}
\newpage

\begin{figure}[]
    \caption{IeRNN training and 1-day ahead forecast on four additional regions.}
    \centering
    \subfloat[Piemonte]{\includegraphics[scale=0.5]{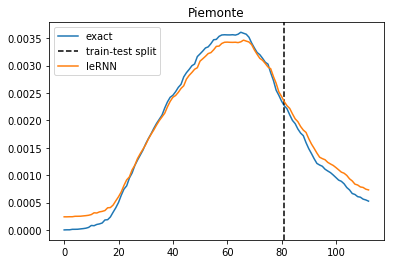}}

    \qquad
    \subfloat[Campania]{\includegraphics[scale=0.5]{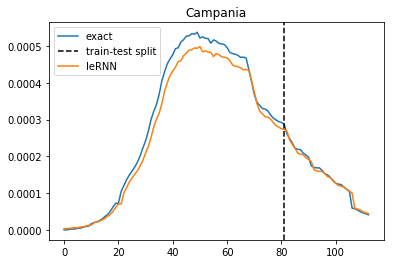}}
    \qquad
    \subfloat[Molise]{\includegraphics[scale=0.5]{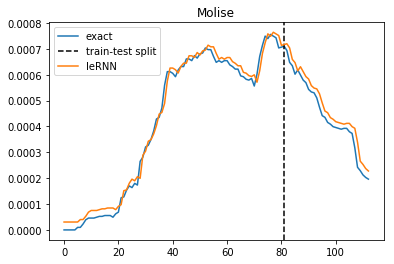}}
    \qquad
    \subfloat[Umbria]{\includegraphics[scale=0.5]{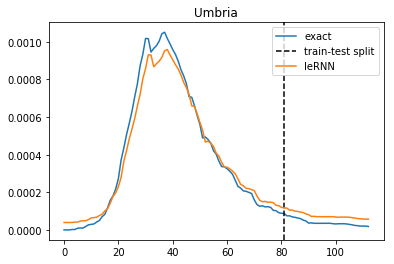}}
    \qquad
    \label{fig:additional}
\end{figure}

\begin{figure}[]
     \caption{Visulization of the latent information $I_e$  in Fig. \ref{fig:additional} learned by IeRNN.}
    \centering
    \subfloat[Piemonte]{\includegraphics[scale=0.5]{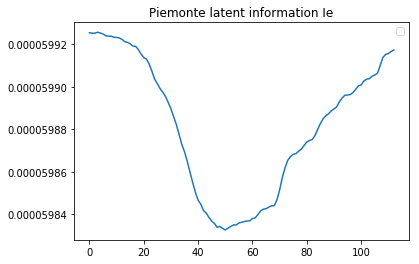}}
    \qquad
    \subfloat[Campania]{\includegraphics[scale=0.5]{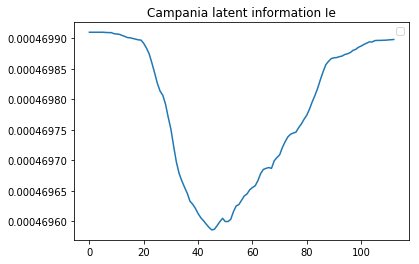}}
    \qquad
    \subfloat[Molise]{\includegraphics[scale=0.5]{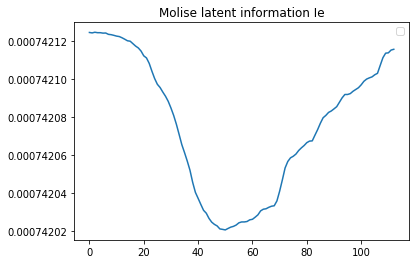}}
    \qquad
    \subfloat[Umbria]{\includegraphics[scale=0.5]{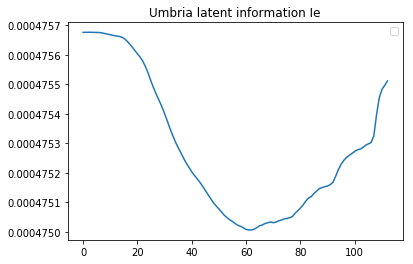}}
   
    \label{fig:Ie}
\end{figure}

\newpage 
\begin{figure*}[ht!]
\caption{Training and 7-day ahead forecast of 3 models (IeRNN, fcNN, and I-model) in 3 rows respectively.}
    \centering
    \subfloat[Lombardy]{\includegraphics[scale=0.45]{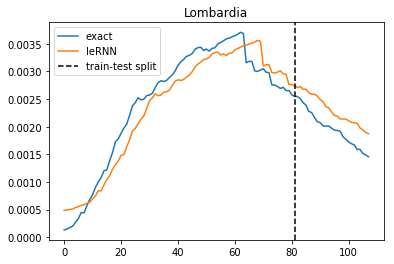}}
    \subfloat[Lazio]{\includegraphics[scale=0.45]{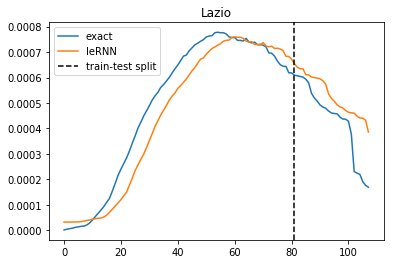}}
    \quad
    \subfloat[Lombardy]{\includegraphics[scale=0.45]{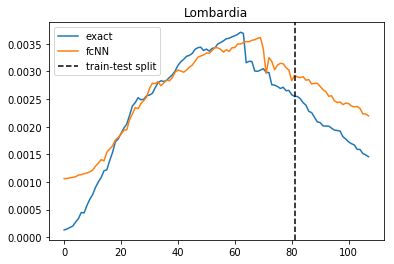}}
    \subfloat[Lazio]{\includegraphics[scale=0.45]{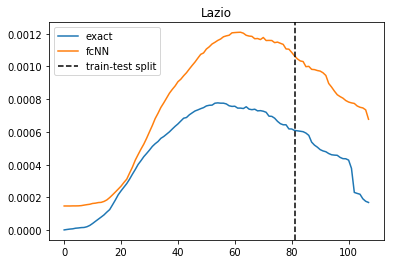}}
    \quad
    \subfloat[Lombardy]{\includegraphics[scale=0.45]{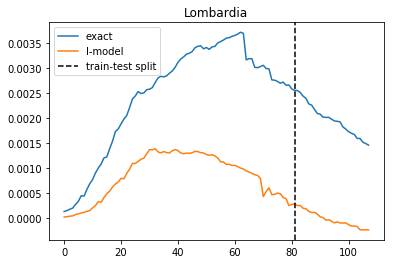}}
    \subfloat[Lazio]{\includegraphics[scale=0.45]{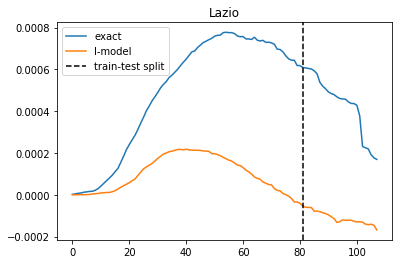}}
    \quad
    
    \label{fig:result 7-day}
\end{figure*}
\begin{figure*}[ht!]
\caption{Training and 7-day ahead forecast of 3 models (IeRNN, fcNN, and I-model) with reduced (40\%) training data in 3 rows respectively.}
    \centering
    \subfloat[Lombardy]{\includegraphics[scale=0.45]{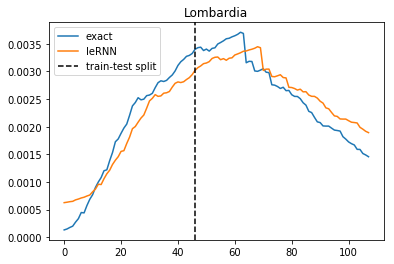}}
    \subfloat[Lazio]{\includegraphics[scale=0.45]{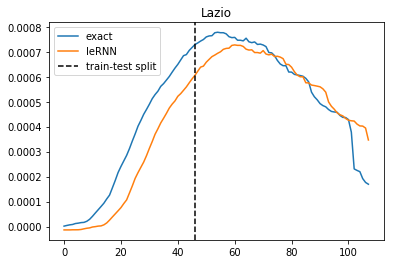}}
    \quad
    \subfloat[Lombardy]{\includegraphics[scale=0.45]{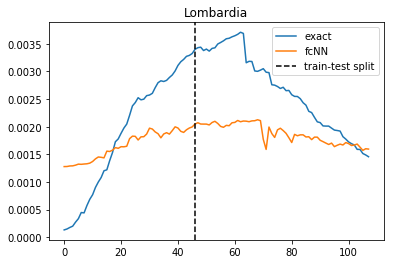}}
    \subfloat[Lazio]{\includegraphics[scale=0.45]{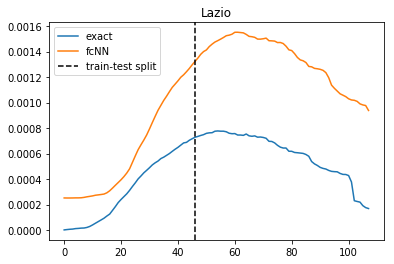}}
    \quad
    \subfloat[Lombardy]{\includegraphics[scale=0.45]{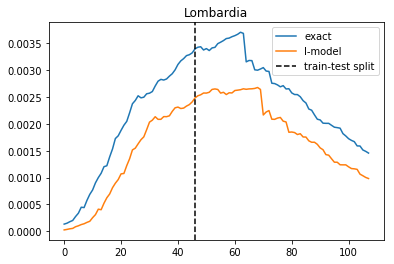}}
    \subfloat[Lazio]{\includegraphics[scale=0.45]{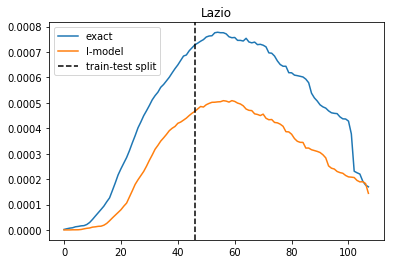}}
    \quad
    
    \label{fig:result2 7-day}
\end{figure*}
\begin{figure*}[ht!]
\caption{Training and 3-day ahead forecast of 3 models (IeRNN, fcNN, and I-model) in 3 rows respectively.}
    \centering
    \subfloat[Lombardy]{\includegraphics[scale=0.45]{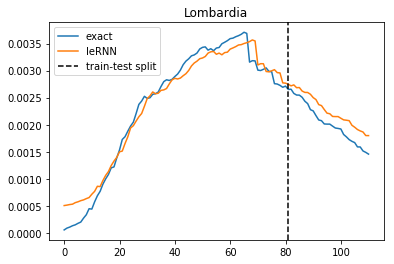}}
    \subfloat[Lazio]{\includegraphics[scale=0.45]{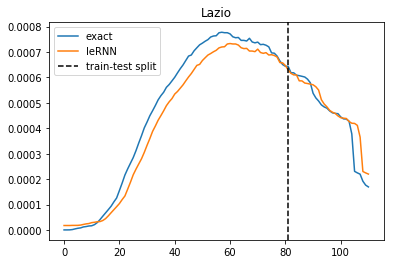}}
    \quad
    \subfloat[Lombardy]{\includegraphics[scale=0.45]{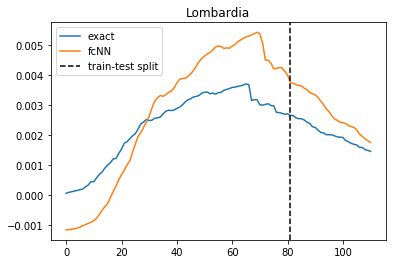}}
    \subfloat[Lazio]{\includegraphics[scale=0.45]{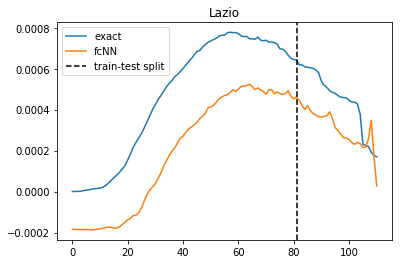}}
    \quad
    \subfloat[Lombardy]{\includegraphics[scale=0.45]{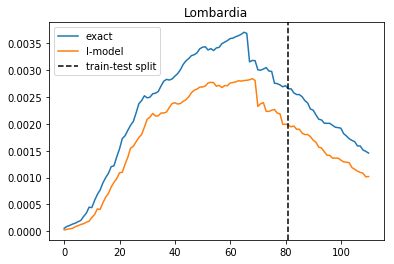}}
    \subfloat[Lazio]{\includegraphics[scale=0.45]{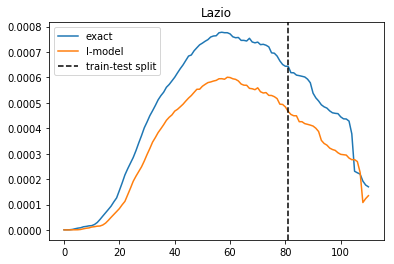}}
    \quad
    
    \label{fig:result 3-day}
\end{figure*}
\begin{figure*}[ht!]
\caption{Training and 3-day ahead forecast of 3 models (IeRNN, fcNN, and I-model) with reduced (40 \%) training data in 3 rows respectively.}
    \centering
    \subfloat[Lombardy]{\includegraphics[scale=0.45]{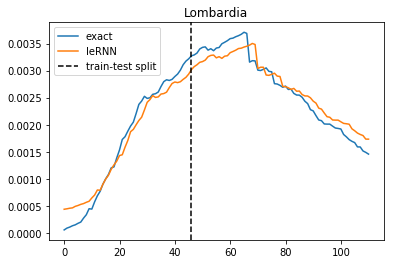}}
    \subfloat[Lazio]{\includegraphics[scale=0.45]{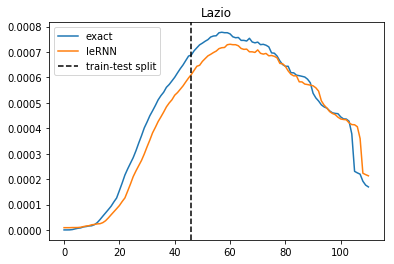}}
    \quad
    \subfloat[Lombardy]{\includegraphics[scale=0.45]{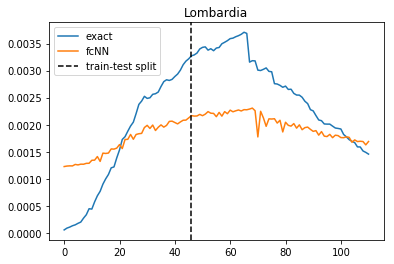}}
    \subfloat[Lazio]{\includegraphics[scale=0.45]{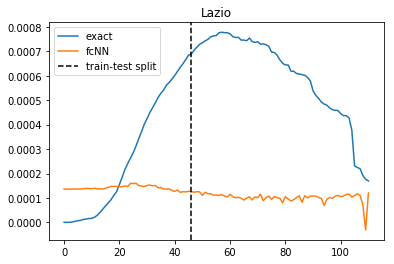}}
    \quad
    \subfloat[Lombardy]{\includegraphics[scale=0.45]{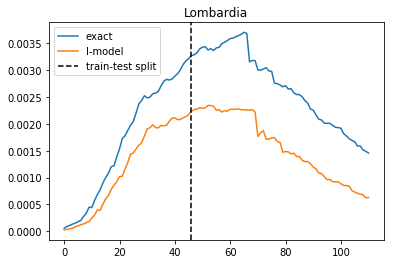}}
    \subfloat[Lazio]{\includegraphics[scale=0.45]{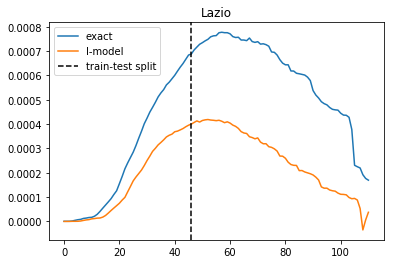}}
    \quad
    
    \label{fig:result 3-day1}
\end{figure*}

\subsection{One-Day Ahead Forecast}
As we see in Fig. \ref{fig:results}, fcNN can perform poorly. This is not a surprise, as both \cite{li_2019} and \cite{yang2015accurate} relied on hundreds of historical observations to train their models. The I-model based on only sequential data in time of one region merely follows the trend of the true data but cannot provide accurate predictions. Our IeRNN model, with the help of additional spatial information, is able to make accurate predictions and outperform other models. We also test the IeRNN with training data reduced to 40\% (46 days). IeRNN is still able to track the general trend of the infected population percentage.
\medskip

We measure the test accuracy with the Root Mean Square Error (RMSE)
averaged over a few trials in training. In Tables \ref{tab:error1a} and \ref{tab:error1b}  on 1-day ahead forecast, IeRNN achieves the smallest RMSE errors, and I-model has the largest errors. The compact I-model with 2 parameters cannot do 1-day ahead prediction as accurately. ARIMA outperforms I-model and does better on Emilia-Romagna and Lazio regions than fcNN. ARIMA, a  linear model, has simpler structure than fcNN whose nonlinearity does not play out in such a short time task. Fig. \ref{fig:additional} shows 
1-day ahead forecast of IeRNN model on other regions with the learned latent external forcing $I_e$ in Fig. \ref{fig:Ie}.
\medskip

\begin{table}[ht!]
    \centering
 \caption{RMSE test errors in 1-day ahead forecast trained with 70 \% of data. E-R= Emilia-Romagna. }
    \begin{tabular}{c c c c}
    \hline \hline
         Model& Lombardy & E-R & Lazio \\
         \hline
         IeRNN & 1.027e-04 & 6.333e-05 & 3.251e-05\\
         \hline
         I-model & 1.175e-03 & 3.284e-04 & 2.439e-04\\
         \hline
         fcNN& 1.580e-04&4.614e-04&2.294e-04\\
         \hline
         ARIMA&9.789e-04&3.627e-04&4.365e-05\\
         \hline
    \end{tabular}
    \label{tab:error1a}
\end{table}

\begin{table}[ht!]
    \centering
    \caption{RMSE test errors in 1-day ahead forecast trained with reduced (40 \% of) data. E-R= Emilia-Romagna. }
    \begin{tabular}{c c c c}
    \hline \hline
         Model& Lombardy & E-R & Lazio \\
         \hline
         IeRNN & 9.850e-05 & 1.778e-04 & 3.617e-05\\
         \hline
         I-model & 1.871e-03 &1.252 e-03 & 5.443e-04\\
         \hline
         fcNN& 3.364e-04&6.204e-04&8.030e-04\\
         \hline
         ARIMA&1.277e-03&1.082e-03&4.018e-05\\
         \hline
         \label{tab:error1b}
    \end{tabular}
    \medskip
    
\end{table}

\begin{table}[ht!]
    \centering
    \caption{RMSE test errors in 7-day ahead forecast trained with 70 \% of data. E-R= Emilia-Romagna.}
    \begin{tabular}{c c c c}
    \hline \hline
         Model& Lombardy & E-R & Lazio \\
         \hline
         IeRNN & 3.513e-04 & 4.423e-04 & 1.161e-04\\
         \hline
         I-model & 2.004e-03 & 6.627e-04 & 5.586e-04\\
         \hline
         fcNN&  6.608e-04&4.804e-04&4.508e-04\\
         \hline
    \end{tabular}
    \medskip
    \label{tab:error2}
\end{table}
\begin{table}[ht!]
    \centering
    \caption{RMSE test errors in 7-day ahead forecast trained with reduced (40 \% of) data. E-R= Emilia-Romagna. }
    \begin{tabular}{c c c c}
    \hline \hline
         Model& Lombardy & E-R & Lazio \\
         \hline
         IeRNN & 3.061e-04 & 4.324e-04 & 7.754e-05\\
         \hline
         I-model & 2.196e-03 & 1.167e-03 & 6.011e-04\\
         \hline
         fcNN& 2.224e-03&6.889e-04&1.851e-04\\
         \hline
    \end{tabular}
    \medskip

    \label{tab:error3}
\end{table}
\begin{table}[ht!]
    \centering
    \caption{RMSE test errors in 3-day ahead forecast trained with 70 \% of data. E-R=  Emilia-Romagna.}
    \begin{tabular}{c c c c}
    \hline \hline
         Model& Lombardy & E-R & Lazio \\
         \hline
         IeRNN & 2.479e-04 & 3.668e-04 & 5.979e-05\\
         \hline
         I-model & 5.609e-04 & 1.724e-04 & 1.383e-04\\
         \hline
         fcNN& 8.165e-04&6.757e-04&1.689e-04\\
         \hline
    \end{tabular}
    \medskip

    \label{tab:error3}
\end{table}

\begin{table}[ht!]
    \centering
    \caption{RMSE test errors in 3-day ahead forecast trained with reduced (40 \% of)  data. E-R= Emilia-Romagna.}
   \begin{tabular}{c c c c}
    \hline \hline
         Model& Lombardy & E-R & Lazio \\
        \hline
         IeRNN & 1.987e-04 & 3.256e-04 & 5.297e-05\\
         \hline
         I-model & 1.114e-03 & 7.337e-04 & 3.507e-04\\
         \hline
         fcNN& 8.611e-04&1.374e-03&5.290e-04\\
        \hline
    \end{tabular}
    \medskip
    
    \label{tab:error3}
\end{table}

\begin{table}[ht!]
    \centering
    \caption{Average model training (tr) and inference (inf) times in seconds on Macbook Pro with Intel i5 CPU. The first two columns are for 70 \% training (tr70) data and the last two columns are for 40 \% training (tr40) data.  }
   \begin{tabular}{c |c| c| c| c}
    \hline \hline
         Model& tr70 &  inf70 &  tr40 & inf40 \\
        \hline
         IeRNN & 0.58s &0.018s & 0.51s &0.02s\\
         \hline
         I-model & 0.14s &0.004s & 0.11s &0.004s\\
         \hline
         fcNN& 0.09s&0.003s &0.09s&0.003s\\
         \hline
         ARIMA&0.23s&0.014s &0.19s&0.015s\\
         \hline
    \end{tabular}
    \medskip
    
    \label{tab:time}
\end{table}




%
\subsection{Multi-Day Ahead Forecast}
In model training for multi-day ahead forecast, 
the training loss function is modified so that the model input comes from multiple days in the past. 
In 7-day ahead forecast, IeRNN leads the other two nonlinear models especially in the 40\% training data case, by as much as a factor of 7 in Lombardy.
In the 3-day ahead forecast, IeRNN leads fcNN by a factor of 4 in the 40\% training data case, as much as a factor of 10 in Lazio.
Figs. 10-13 show model comparison in training and forecast phases for Lombardy and Lazio.

\subsection{Model Size and Computing Time}
IeRNN (fcNN) has about 16400 (1800) parameters.  
The optimized  $(\beta_1,\beta_2, \gamma)=  
(0.685,0.158,0.044)$ in Lombardy, similarly in other regions. 
 Table \ref{tab:time} 
 lists average model training and inference times.


%

\section{Conclusions and Future Work}
We developed a novel spatiotemporal infectious disease model consisting of a discrete epidemic equation for the region of interest and RNNs for interactions with nearest geographic regions. Our model can be trained under 1 second. Its inference takes a fraction of a second, suitable for real-time applications. Our model out-performs temporal models in one-day and multi-day ahead forecasts in  limited training data regime.  In future work, we shall consider social and control mechanisms \cite{Pareschi_2020,Levin_2020} to strengthen the I-equation, as well as  traffic data to expand interaction beyond nearest neighbors. 

\section{Acknowledgements}

\noindent The work was partially supported by NSF grants IIS-1632935, and  DMS-1924548. JX would like to thank Prof. Fred Wan for helpful communications on disease modeling.

\newpage

\bibliographystyle{plain}
\bibliography{main}

\begin{thebibliography}{10}

\bibitem{Pareschi_2020}
G.~Albi, L.~Pareschi, and M.~Zanella.
\newblock Control with uncertain data of socially structured compartmental
  epidemic models.
\newblock {\em arXiv preprint arXiv}, 2004.13067v1:1--26, 2020.

\bibitem{anderson_may_1992}
R.~Anderson and R.~May.
\newblock {\em Infectious Diseases of Humans: Dynamics and Control}.
\newblock Oxford University Press, Oxford, 1992.

\bibitem{hethcote_2000}
H.~Hethcote.
\newblock The mathematics of infectious diseases.
\newblock {\em SIAM Review}, 42:599 -- 653, 2000.

\bibitem{hochreiter1997long}
S.~Hochreiter and J.~Schmidhuber.
\newblock Long short-term memory.
\newblock {\em Neural computation}, 9(8):1735--1780, 1997.

\bibitem{Italy_data}
{I}talian region.
\newblock {\em Italian COVID-19 Data (accessed March-July, 2020)}.
\newblock
  https://github.com/Akaza994/COVID-19-Data/blob/master/covid-19/italy.csv.

\bibitem{li_2019}
Z.~Li, X.~Luo, B.~Wang, A.~Bertozzi, and J.~Xin.
\newblock A study on graph-structured recurrent neural networks and
  sparsification with application to epidemic forecasting.
\newblock In {\em World Congress on Global Optimization}, pages 730--739.
  Springer, 2019, https://doi.org/10.1007/978-3-030-21803-4\textunderscore 80.

\bibitem{Levin_2020}
D.~Morris, F~Rossine, J.~Plotkin, and S.~Levin.
\newblock Optimal, near-optimal, and robust epidemic control.
\newblock {\em arXiv preprint arXiv}, 2004.02209v2:1--28, 2020.

\bibitem{roosa_2020}
K.~Roosa, Y.~Lee, R.~Luo, A.~Kirpich, R.~Rothenberg, J.M. Hyman, P.~Yan, and
  G.~Chowell.
\newblock Real-time forecasts of the {COVID}-19 epidemic in {China} from
  {February} 5th to {February} 24th, 2020.
\newblock {\em Infectious Disease Modelling}, 5:256 -- 263, 2020.

\bibitem{wang2018graph}
B.~Wang, X.~Luo, F.~Zhang, B.~Yuan, A.~Bertozzi, and P.~Brantingham.
\newblock Graph-based deep modeling and real time forecasting of sparse
  spatio-temporal data.
\newblock {\em arXiv preprint arXiv:1804.00684}, 2018.

\bibitem{wang}
B.~Wang, P.~Yin, A.~Bertozzi, P.~Brantingham, S.~Osher, and J.~Xin.
\newblock Deep learning for real-time crime forecasting and its ternarization.
\newblock {\em Chinese Annals of Mathematics, Series B}, 40(6):949--966, 2019.

\bibitem{yang2015accurate}
S.~Yang, M.~Santillana, and S.~Kou.
\newblock Accurate estimation of influenza epidemics using {G}oogle search data
  via {ARGO}.
\newblock {\em Proceedings of the National Academy of Sciences},
  112(47):14473--14478, 2015.

\end{thebibliography}

\newpage
\end{document}